\documentclass[a4paper,12pt]{article}
\usepackage{amsmath, amsthm, amsfonts, amssymb}

\numberwithin{equation}{section}

\begin{document}

\begin{center}
{\bf \large On regular and singular perturbations of acoustic and
quantum waveguides}
\end{center}

\medskip
\begin{center}
Rustem R. GADYL'SHIN\footnote{The work is supported by grants of
RFBR (02-01-00693, 02-01-768) and by the program ''Scientific
Schools'' (1446.2003.1).}
\end{center}

\medskip

\begin{quote}
{ \emph{Bashkir State Pedagogical University, October Revolution
St.~3a, 450000, Ufa, Russia\\
Institute of Mathematics,  Ufa
Science Center of Russian Academy of Sciences, 112 Chernyshevski
str., 450077 Ufa, Russia
\\ E-mail:} \texttt{gadylshin@bspu.ru, gadylshin@narod.ru} }
\end{quote}

\medskip

\begin{center}
Abstract
\end{center}
\begin{quote}\quad
{\small We consider regular and singular perturbations of the
Dirichlet and Neumann boundary value problems for the Helmholtz
equation in $n$-dimensional cylinders. Existence of eigenvalues
and their asymptotics are studied.}
\end{quote}

\bigskip

\section{Introduction}

We consider regular and singular perturbations of the Neumann and
Dirichlet boundary value problems for
$\mathcal{H}^{(m)}_0:=-(\Delta+\mu_m)$ in $n$-dimensional cylinder
$\Pi=(-\infty,\infty)\times\Omega$, where
$\Omega\subset\mathbb{R}^{n-1}$ is a simply connected bounded
domain with $C^\infty$-boundary for $n\ge3$ and is an interval
$(a,b)$ for $n=2$. Hereinafter, $\mu_j$ and $\phi_j$ are the
eigenvalues and eigenfunctions of
$-\Delta':=-\left(\frac{\partial^2}{\partial
x_2^2}+\dots+\frac{\partial^2}{\partial x_n^2}\right)$ in $\Omega$
subject to the same type of the boundary condition on
$\partial\Omega$ as in the original unperturbed  boundary valued
problem for $\mathcal{H}^{(m)}_0:=-(\Delta+\mu_m)$ on $\partial
\Pi$, $\mu_j<\mu_{j+1}$, $j=1,2,\dots$. The functions $\phi_j$ are
assumed to be normalized in  $L^2(\Omega)$. The Neumann problem is
a mathematical model describing acoustic waveguide while the
Dirichlet one corresponds to a quantum waveguide. It is known that
unperturbed boundary value problems have no eigenfunctions in
$H^1(\Pi)$. At the same time such eigenfunctions and eigenvalues
(bound states) can emerge under perturbations. We study the
questions on existence and absence of such emerging eigenvalues
and constructing their asymptotic expansions. Both cases of
regular and singular perturbations of these boundary value
problems are considered. The regular perturbation treated in the
next section is performed by a small localized linear operator of
second order. The example of such operator is a small complex
potential as well as other perturbations considered in \cite{G4}
for the Schr\"odinger operator on the axis. Other examples are
small deformations of strips and cylinders which can be reduced to
the case we consider by a change of variables
\cite{DE}--\cite{BEGK}. As a singular perturbation of the
Dirichlet and Neumann boundary value problems in $\Pi$ we consider
the switching of type of boundary condition at a small segment of
the boundary. Such a choice is motivated by a number of articles
having appeared recently and containing both rigorous results for
quantum waveguides (\cite{BGRS}, \cite{EV1}) as well as
non-rigorous asymptotic results (see \cite{Po}, \cite{8} and other
articles of these authors on singularly perturbed two- and
three-dimensional quantum waveguides given in the bibliography of
\cite{Po}, \cite{8}). These formal asymptotics were derived by the
method of matching of asymptotic expansions \cite{I} on the basis
of scheme employed in \cite{G2}--\cite{G3} for constructing the
asymptotics for scattering frequencies of Helmholtz resonator.
However, rigorous justification of the asymptotics for these
scattering frequencies adduced in \cite{G2}--\cite{G3} is based on
the compactness of obstacle (boundary) and due to this fact it can
not be applied to the case of a waveguide. The question on an
estimating of the inverse operator for singularly perturbed
waveguides (providing a possibility to justify formal asymptotics)
is treated in the third section. In two concluding sections we
construct the leading terms for asymptotics of the eigenvalues and
poles for singularly perturbed quantum and acoustic waveguides.

\section{Regular perturbation}

Hereinafter $H_{loc}^j({\Pi})$ is a set of functions defined on
${\Pi}$ whose restriction to any bounded domain $D\subset {\Pi}$
belongs to $H^j(D)$, $\|\bullet\|_G$ and $\|\bullet\|_{j,G}$ are
norms in $L^2(G)$ and $H^j(G)$, respectively. Next, let
$Q=(-R,R)\times\Omega$, where $R>0$ is an arbitrary fixed number,
$L^2({\Pi};Q)$ be the subset of functions in $L^2({\Pi})$ with
supports in $\overline {Q}$, ${\mathcal L}_\varepsilon$ be linear
operators mapping $H_{loc}^2({\Pi})$ into $L^2({\Pi};Q)$ such that
$\|{\mathcal L}_\varepsilon[u]\|_{Q}\le C({\mathcal
L})\,\|u\|_{2,Q}$, where constant $C({\mathcal L})$ is independent
of $\varepsilon$, $0<\varepsilon\ll1$. In this section we study
the existence and the asymptotics of the eigenvalues of the
Dirichlet and Neumann boundary value problems for
$\mathcal{H}^{(m)}_\varepsilon:=\mathcal{H}^{(m)}_0-\varepsilon{\mathcal
L}_\varepsilon$ in $\Pi$. For a small complex $k$, we define a
linear operator $A^{(m)}(k)\,:\,L^2({\Pi};Q)\to H^2_{loc}({\Pi})$
as
\begin{equation}\label{A}
A^{(m)}(k)g:=\left(
\sum\limits_{j=1}^{m-1}+\sum\limits_{j=m}^\infty\right)\frac{\phi_j(x')}{2K_j^{(m)}(k)}\int\limits_{\Pi}
e^{-K_j^{(m)}(k)|x_1-t_1|}\phi_j(t') g(t)\,dt,
\end{equation}
where $x'=(x_2,...,x_n)$,
$K_j^{(m)}(k)=\mathrm{i}\sqrt{\mu_m-\mu_j-k^2}$ for $j<m$,
$K_m^{(m)}(k)=k$ and $K_j^{(m)}(k)=\sqrt{\mu_j-\mu_m+k^2}$ for
$j>m$. By analogy with \cite{G4} for $f\in L^2(\Pi;Q)$ we seek a
solution of
\begin{align} & \mathcal{H}^{(m)}_\varepsilon u_\varepsilon=-k^2u_\varepsilon+f
\quad \hbox{as $x\in\Pi$}, \quad
u_\varepsilon=0\quad\left(\hbox{or $\frac{\partial
u_\varepsilon}{\partial\nu}=0$}\right)\quad\hbox{as
$x\in\partial\Pi$}\label{2N}
\end{align}
(where $\nu$ is normal) as
\begin{equation}
u_\varepsilon=A^{(m)}(k)g_\varepsilon,\label{g}
\end{equation}
where  $g_\varepsilon\in L^2({\Pi};Q)$. By definition (\ref{g}) is
the solution of the equation
$\mathcal{H}^{(m)}_0(k)u_\varepsilon=-k^2u_\varepsilon+g_\varepsilon$
in $\Pi$ and satisfies the boundary condition in (\ref{2N}).
Substituting (\ref{g}) into (\ref{2N}), we get that (\ref{g})
gives a solution for (\ref{2N}) if
\begin{equation} (I-\varepsilon
{\mathcal L}_\varepsilon A^{(m)}(k))g_\varepsilon=f,\label{F1}
\end{equation}
where $I$ is identity mapping. If
$\mathcal{L}_\varepsilon[\phi_m]=0$,  due to (\ref{A}), (\ref{g})
and (\ref{F1}) it follows that the pole $k_\varepsilon^{(m)}$ of
(\ref{g}) is equal zero and $g_\varepsilon\to0$ as
$\varepsilon\to0$. Thus, there is no small eigenvalue in this
case. Assume $\mathcal{L}_\varepsilon[\phi_m]\not=0$,
\begin{equation*}\label{S}
\begin{aligned}
\left<F\right>:=\int\limits_{\Pi}F\,dx,\quad\widetilde{T}^{(m)}_\varepsilon(k)g:=
&{\mathcal
L}_\varepsilon[A^{(m)}(k)g]-\frac{\left<g\phi_m\right>}{2k}
{\mathcal L}_\varepsilon[\phi_m],\\
S^{(m)}_\varepsilon(k):=&\left(I-\varepsilon
\widetilde{T}^{(m)}_\varepsilon(k)\right)^{-1}.
\end{aligned}
\end{equation*}
Applying the operator $S^{(m)}_\varepsilon(k)$ to both sides of
the equation (\ref{F1}), we obtain that \begin{align}
&\left(g_\varepsilon-\varepsilon\frac{\left<g_\varepsilon\phi_m\right>}{2k}S^{(m)}_\varepsilon(k){\mathcal
L}_\varepsilon[\phi_m]\right)=
S^{(m)}_\varepsilon(k)f,\label{F2}\\
&\left<g_\varepsilon\phi_m\right>\left(1-\frac{\varepsilon}{2k}
\left<\phi_mS^{(m)}_\varepsilon(k){\mathcal
L}_\varepsilon[\phi_m]\right>
\right)=\left<\phi_mS^{(m)}_\varepsilon(k)f\right>.\label{F3}
\end{align}
 The equality
(\ref{F3}) allows us to determine
$\left<g_\varepsilon\phi_m\right>$. Substituting its value into
(\ref{F2}), we easily get the formula
\begin{equation}
g_\varepsilon=\varepsilon\frac{2k\left<S^{(m)}_\varepsilon(k)f\right>
S^{(m)}_\varepsilon(k)\mathcal{L}_\varepsilon[\phi_m]}
{2k-\varepsilon\left<\phi_mS^{(m)}_\varepsilon(k)\mathcal{L}_\varepsilon[\phi_m]\right>}+
S^{(m)}_\varepsilon(k)f.\label{G1}
\end{equation}
Formulas (\ref{G1}) and (\ref{g}) imply, that, if
$k_\varepsilon^{(m)}$ is a solution of the equation
\begin{equation}
{2k-\varepsilon\left<\phi_mS^{(m)}_\varepsilon(k)\mathcal{L}_\varepsilon[\phi_m]\right>}=0,
\label{E1}
\end{equation}
then the residue  of (\ref{g}) at $k_\varepsilon^{(m)}$:
\begin{equation}
\psi_\varepsilon^{(m)}=A^{(m)}(k_\varepsilon^{(m)})S^{(m)}_\varepsilon(k_\varepsilon^{(m)}){\mathcal
L}_\varepsilon[\phi_m]\label{E2}
\end{equation}
is the solution of the equation
$\mathcal{H}^{(m)}_\varepsilon\psi_\varepsilon^{(m)}=\lambda_\varepsilon^{(m)}\psi_\varepsilon^{(m)}$
in $\Pi$ (with corresponding homogeneous Dirichlet or Neumann
boundary conditions), where
$\lambda_\varepsilon^{(m)}=-\left(k^{(m)}_\varepsilon\right)^2$.
The formulas (\ref{A}), (\ref{E2}) yield if
$\mathrm{Re}\,k_\varepsilon^{(1)}>0$, then
$\psi_\varepsilon^{(1)}\in L^2(\Pi)$ and, hence,
$\lambda_\varepsilon^{(1)}$ is the eigenvalue which due to
(\ref{E1}) has the asymptotics
\begin{equation}\label{as}
\lambda_\varepsilon^{(m)}=-\varepsilon^2\frac{1}{4}
\left<\phi_m\mathcal{L}_\varepsilon[\phi_m]\right>^2+
O\left(\varepsilon^3\right)
\end{equation}
with $m=1$ (and the function (\ref{E2}) is the associated
eigenfunction). For $m\ge2$, the formulas (\ref{A}), (\ref{E1}),
(\ref{E2}) imply, that if $\mathrm{Re}\,k_\varepsilon^{(m)}>0$ and
$\mathrm{Im}\,k_\varepsilon^{(m)}>0$, then
$\psi_\varepsilon^{(m)}\in L^2(\Pi)$, too, and, hence,
$\lambda_\varepsilon^{(m)}$ is the eigenvalue of the perturbed
problem  with asymptotics (\ref{as}). In particular, the equation
(\ref{E1}) allows us to maintain that in the case
$\left<\phi_1\mathcal{L}_\varepsilon[\phi_1]\right>\ge\delta>0$
there exists small eigenvalue.

\section{Singular perturbations.
Convergence of poles and representation of solutions near poles}

Assume for simplicity in describing the of perturbations that the
domain $\Omega$ coincides with half-space $x_n>0$ in some
neighborhood of the origin (in variables $x'$), $\omega$ is a
$(n-1)$-dimensional bounded domain in the hyperplane $x_n=0$
having  smooth boundary, $\omega_\varepsilon=\{x:\,
x\varepsilon^{-1}\in\omega\}$,
$\Gamma_\varepsilon=\partial\Pi\backslash\overline{\omega_\varepsilon}$.
For a given $f\in L^2({\Pi};Q)$, we consider two singularly
perturbed boundary value problems
\begin{equation}\begin{aligned} &
\mathcal{H}_0^{(m)}u_\varepsilon=-k^2u_\varepsilon+f,\,\,\,
x\in\Pi, \\ &u_\varepsilon=0,\,\,\,
x\in\Gamma_\varepsilon\,\,(\hbox{or
$x\in\omega_\varepsilon$}),\quad \frac{\partial
u_\varepsilon}{\partial\nu}=0,\,\,\,
x\in\omega_\varepsilon\,\,(\hbox{or $x\in\Gamma_\varepsilon$}).
\end{aligned}\label{SD}
\end{equation}

Let $\Gamma^{R}_0=\partial\Pi\cap\partial Q$, $\Omega^R=\partial
Q\backslash\overline{\Gamma^{R}_0}$,
$\Gamma^{R}_\varepsilon=\Gamma^{R}\backslash
\overline{\omega_\varepsilon}$. For each $V\in H^2(Q)$, we denote
by $\sigma_\varepsilon:\, H^2(Q)\to H^1(Q)$ the inverse operator
for the following boundary value problems
\begin{equation*}
\begin{aligned}
\Delta W_\varepsilon &= \Delta V\,, \quad x\in Q,\qquad
W_\varepsilon=V,\quad x\in \Omega^R,
\\
W_\varepsilon&=0,\quad x\in\Gamma^R_\varepsilon\,\,(\hbox{or
$x\in\omega_\varepsilon$}),\qquad \frac{\partial
W_\varepsilon}{\partial\nu}=0,\quad
x\in\omega_\varepsilon\,\,(\hbox{or $x\in\Gamma^R_\varepsilon$}).
\end{aligned}
\end{equation*}
Let $\chi^\pm(x_1)$ be an infinitely differentiable mollifier
function equalling to one for $\pm x_1\le R/2$ and vanishing for
$\pm x_1\ge R$, $\Pi_\pm=\{x:\,x\in\Pi,\,\pm x_1>0\}$, $p_\pm$ be
the restriction operator from $\Pi$ to $\Pi_\pm$, $p^Q_\pm$ be the
restriction operator from $\Pi_\pm$ to $\Pi_\pm\cap Q$,
\begin{equation*}
 \begin{aligned}
 A^{(m)}_\pm(k)g^\pm:=&
\sum\limits_{j=1}^\infty\frac{\phi_j(x')}{2K_j^{(m)}(k)}
\int\limits_{\Pi_\pm}
\Big(e^{-K_j^{(m)}(k)|x_1-t_1|}
\\ &\qquad\qquad\qquad\qquad-e^{-K_j^{(m)}(k)|x_1+t_1|}\Big)\phi_j(t')
g^\pm(t)\,dt,\quad x\in\Pi_\pm, \\
\mathcal{A}^{(m)}_\varepsilon(k)g:=& (1-\chi^+)A^{(m)}_+(k)p_+g+
(1-\chi^-)A^{(m)}_-(k)p_-g\\ & +\chi^+\chi^-\sigma_\varepsilon
\left(p_+^QA^{(m)}_+(k)p_+g+ p_-^QA^{(m)}_-(k)p_-g\right),\quad
g\in L^2(\Pi;Q).
\end{aligned}
\end{equation*}
We construct the solution of (\ref{SD}) in  the form
\begin{align}\label{SR}
u_\varepsilon=\mathcal{A}^{(m)}_\varepsilon(k)g_\varepsilon,
\end{align}
where $g_\varepsilon$ is a some function belonging to
$L^2({\Pi};Q)$. Substituting (\ref{SR}) into (\ref{SD}), by
analogy with \cite{1+1} we deduce that this function is a solution
of (\ref{SD}) in the case
\begin{equation}
g_\varepsilon=(I+T_\varepsilon^{(m)} (k))^{-1}f,\label{SF1}
\end{equation}
where, for any fixed $\varepsilon$,  $T_\varepsilon^{(m)}(k)$ is a
holomorphic operator-valued function and, for any fixed $k$,
$T_\varepsilon^{(m)}(k)$ is a compact operator in $L^2(\Pi;Q)$.
Analyze of this family with respect to $\varepsilon$ (which is
similar to \cite{G1} and based on \cite{1+1}) and the
representations (\ref{SR}),  (\ref{SF1}) imply that there exists
one pole $k_\varepsilon^{(m)}\to0$ of the solution of (\ref{SD})
and for small $k$, this solution meet the representation

\begin{gather}
u_\varepsilon(x,k) =\frac{\psi^{(m)}_{\varepsilon}(x)}{
2\left(k-k_\varepsilon^{(m)}\right)}\int\limits_{\Pi}\psi^{(m)}_{\varepsilon}(y)\,f(y)\,dy+
\widetilde u_\varepsilon(x,k),\\
\|\widetilde{u}_\varepsilon\|_{1,D}\le C(D,Q)\|f\|_\Pi\label{EST}
\end{gather}
for any bounded domain $D\subset\Pi$. The residue
$\psi_\varepsilon^{(m)}$ at this pole is a solution  to the
equation
$\mathcal{H}_0^{(m)}\psi_\varepsilon^{(m)}=\lambda_\varepsilon^{(m)}
\psi_\varepsilon^{(m)}$ in $\Pi$, where
$\lambda^{(m)}_\varepsilon=-\left(k_\varepsilon^{(m)}\right)^2$,
satisfies the boundary conditions from (\ref{SD}) and for any
fixed $x_1$ converges to $\phi_m$ (up to a multiplicative
constant) as $\varepsilon\to0$. This convergence, the
representation (\ref{SR}) and  the definition of
$\mathcal{A}_\varepsilon^{(m)}(k)$ imply that
\begin{equation*}
\begin{aligned}
\psi^{(m)}_{\varepsilon}(x)=
\sum\limits_{j=1}^{m-1}a_j^\varepsilon\phi_j(x')
e^{-|x_1|\mathrm{Re}\,K_j^{(m)}(k_\varepsilon^{m})}+&a_m^\varepsilon\phi_m(x')
e^{-|x_1|\mathrm{Re}\,k_\varepsilon^{(m)}}
+o\left(e^{-|x_1|\delta}\right)\\&\hbox{as $|x_1|\to\infty$}
\end{aligned}
\end{equation*}
where $a_m^\varepsilon=1+o(1)$ as $\varepsilon\to0$ and $\delta>0$
some fixed number. In partially,  this asymptotics implies that
\begin{align}
&\hbox{there exists eigenvalue $\lambda^{(1)}_\varepsilon$
provided $\mathrm{Re}\,k_\varepsilon^{(1)}>0$},\label{a}\\
&\begin{aligned} &\hbox{if $m\ge2$,
$\mathrm{Re}\,k_\varepsilon^{(m)}>0$ but
$\mathrm{Im}\,k^{(m)}_\varepsilon<0$ and
$a_1^\varepsilon\not=0$,}\\ &\quad\hbox{then there is no an
eigenvalue},\end{aligned}\label{c}\\ &\hbox{there is no an
eigenvalue if $\mathrm{Re}\,k_\varepsilon^{(m)}\le0$}.\label{b}
\end{align}

Thus, in fact we need to construct and to justify asymptotics of
the pole $k_\varepsilon^{(m)}$ (and, an additional, asymptotics of
the residue $\psi_\varepsilon^{(m)}$ in the case (\ref{c})) which
generates the eigenvalue or doesn't. As above mentioned in the
case of regular perturbation the asymptotics for pole can obtained
by simple calculations in (\ref{E1}), while dealing with singular
perturbation, we have no such equation. On the other hand, the
representation (\ref{EST}) allows to justify the method of
matching asymptotic expansions in constructing the asymptotics for
the poles $k_\varepsilon^{(m)}$ and for the residue
$\psi_\varepsilon^{(m)}$.

As it has been mentioned above, the formal construction of
complete asymptotics of poles for the boundary valued problems
(\ref{SD}) and for Helmholtz resonator \cite{G2}--\cite{G3} is
similar. That's why in the next two section we will construct
first perturbed terms of poles only.

\section
{Singular perturbation of quantum waveguide. Asymptotics of poles
and eigenvalues}

Let $S_n$ be the unit sphere in $\mathbb{R}^n$,
$G^{(\mathcal{D})}_m(x,y,k)$ be the Green function of the
unperturbed Dirichlet boundary value problem in $\Pi$,
${\Phi}_m=\frac{\partial}{\partial x_n}\phi_m(x')|_{x'=0}\not=0$,
$\Psi_m^{(\mathcal{D})}(x,k)=-2k{\Phi}_m^{-1}\frac{\partial}{\partial
y_n}G^{(\mathcal{D})}_m(x,y,k)|_{y=0}$. By definition
\begin{align}&\Psi_m^{(\mathcal{D})}(x,k)\to\phi_m(x'),\qquad
k\to0\quad\hbox{for any fixed $x\not=0$}\label{D4} ,\\ \label{D3}
&\Psi_m^{(\mathcal{D})}(x,k)={\Phi}_mx_n+\frac{4k}{{\Phi}_m|S_n|}\frac{x_n}{r^n}+O\left(kr^{-n+2}\right),\qquad
r=|x|\to0,\,k\to0.
\end{align}
Taking into account (\ref{D4}), outside small neighborhood of
$\omega_\varepsilon$ we construct the residue
$\psi_\varepsilon^{(m)}$  in the form
$\psi_\varepsilon^{(m)}(x)\sim\Psi_m^{(\mathcal{D})}(x,k_\varepsilon^{(m)})$.
Near $\omega_\varepsilon$ we construct asymptotics by using the
method of matching asymptotic expansions \cite{I}--\cite{G3} in
the variables $\xi=\varepsilon^{-1}x$. The structure of the
expansions of $\psi_\varepsilon^{(m)}$ in this zone and of the
pole $k_\varepsilon^{(m)}$ are inspired by the following
consideration. When $x=\varepsilon\xi$ and
$k=k_\varepsilon^{(m)}$, both terms in right hand side of
(\ref{D3}) must have the same order with respect to $\varepsilon$.
This degree determines the first term in the interior layer for
$\psi_\varepsilon^{(m)}$, while the right hand side of (\ref{D3})
(rewritten in variables $\xi$ and for $k=k_\varepsilon^{(m)}$)
determines the asymptotics of this term as $\rho=|\xi|\to\infty$.
Due to these reasons we construct asymptotics as
\begin{align}\label{D5}
k_\varepsilon^{(m)}&=\varepsilon^n\tau^{(m)}_n+\dots,\quad
\psi_\varepsilon^{(m)}(x)=\varepsilon v^{(m)}_1(\xi)+\dots,\\
v^{(m)}_1(\xi)&=\Phi_m\xi_n+4\tau^{(m)}_n
\left({\Phi}_m|S_n|\right)^{-1}\xi_n\rho^{-n}+o\left(\rho^{-n+1}\right),\qquad
\rho\to\infty.\label{D7}
\end{align}
Substituting (\ref{D5}) in (\ref{SD}) (with $f=0$ and
$k=k_\varepsilon^{(m)}$), we obtain the boundary value problem for
$v^{(m)}_1$:
\begin{align} & \Delta_\xi v^{(m)}_1=0,\quad \xi_n>0,
\qquad v^{(m)}_1=0,\quad \xi\in\Gamma,\qquad \frac{\partial
v^{(m)}_1}{\partial\xi_n}=0,\quad \xi\in\omega,\label{SD1}
\end{align}
where $\Gamma=\{\xi:\,\xi_n=0,\,\xi\notin\omega\}$. It is known,
there exists the solution $X_n$ of (\ref{SD1}) with asymptotics
$X_n(\xi)=\xi_n+c_n(\omega)\xi_n\rho^{-n}+o\left(\rho^{-n+1}\right)$
as $\rho\to\infty$, where $c_n(\omega)>0$. Thus it follows from
(\ref{D7}) that
\begin{equation}\label{Dir}
v^{(m)}_1(\xi)=\Phi_mX_n(\xi),\qquad\tau^{(m)}_n=
4^{-1}c_n(\omega)|S_n|\Phi^2_m>0.
\end{equation}
By (\ref{D5}), (\ref{Dir}) we have
$\mathrm{Re}\,k_\varepsilon^{(m)}>0$ and, hence (see (\ref{a})),
there exists eigenvalue
\begin{equation*}\label{Dir1}
\lambda_\varepsilon^{(1)}=-\varepsilon^{2n}\left(\frac{c_n(\omega)|S_n|
\Phi^2_1}{4}\right)^2+o\left(\varepsilon^{2n}\right).
\end{equation*}

For $m\ge2$, constructing next terms for expansions
$k_\varepsilon^{(m)}$ and $\psi_\varepsilon^{(m)}$  (similar
\cite{G2}--\cite{G3}) one can obtain that
\begin{equation*}\label{Dir11}\begin{aligned}
\mathrm{Im}\,k^{(m)}_\varepsilon
=&-\varepsilon^{2n}\left(\frac{c_n(\omega)|S_n|\Phi_m}{4}\right)^2\,\,
\sum\limits_{j=1}^{m-1}\frac{\Phi_j^2}{\sqrt{\mu_m-\mu_j}}+
o\left(\varepsilon^{2n}\right)<0,\\ a^\varepsilon_1
\sim&\frac{k_\varepsilon^{(m)}\Phi_1}
{K_1^{(m)}(k_\varepsilon^{(m)})\Phi_m}\not=0.\end{aligned}
\end{equation*}
where ${\Phi}_j=\frac{\partial}{\partial x_n}\phi_j(x')|_{x'=0}$.
Therefore, the pole $k_\varepsilon^{(m)}$ meets the asymptotics
(\ref{D5}), (\ref{Dir}), but  (see (\ref{c})) does not generate an
eigenvalue of the considered singular perturbation of the
Dirichlet boundary value problem.

\section{Singular perturbation of acoustic waveguide. Asymptotics of poles}

Let $G^{(\mathcal{N})}_m(x,y,k)$ be the Green function of the
unperturbed Neumann boundary value problem, $\phi_m(0)\not=0$,
$\Psi_m^{(\mathcal{N})}(x,k)=-2k\phi_m^{-1}(0)G_m^{(\mathcal{N})}(x,0,k)$,
$\alpha_n(r)=r^{-n+2}$ for $n\ge 3$ and $\alpha_2(r)=-\ln r$. By
definition
\begin{equation}
\begin{aligned}
&\Psi_m^{(\mathcal{N})}(x,k)\to\phi_m(x'),\qquad
k\to0\quad\hbox{for any fixed $x\not=0$},\\
&\Psi_m^{(\mathcal{N})}(x,k)={\phi}_m(0)+4k\left(\phi_m(0)|S_n|\right)^{-1}
\alpha_n(r)+O\left(kr^{-n+3-\delta^2_n}\right),\\ &\hskip8cm
r\to0,\,k\to0,
\end{aligned}\label{N4}
\end{equation}
where $\delta_J^s$ is the Cronecker delta.  Taking into account
(\ref{N4}) and following the method of matching asymptotic
expansions similar the previous section  we obtain that
\begin{equation}\label{NA}
\begin{aligned}
&k_\varepsilon^{(m)}=\varepsilon^{n-2}\tau^{(m)}_{n-2}+\dots,\quad
n\ge3,\qquad
k_\varepsilon^{(m)}=-\ln^{-1}\varepsilon\,\tau^{(m)}_0+\dots,\quad
n=2,\\
&\tau^{(m)}_{n-2}=-\frac{C_n(\omega)|S_n|\phi^2_m(0)}{4}<0,\quad
n\ge3,\quad\tau^{(m)}_0=-\frac{\pi \phi^2_m(0)}{2}<0, \quad n=2,
\end{aligned}
\end{equation}
where $C_n(\omega)>0$ is the capacity of the disk $\omega$. Thus,
$\mathrm{Re}\,k_\varepsilon^{(m)}<0$. Therefore, the pole
$k_\varepsilon^{(m)}$ meets the asymptotics (\ref{NA}), but (see
(\ref{b})) it does not generate an eigenvalue of the considered
singular perturbation of the Neumann boundary value problem.

\bigskip

 The
author thanks E.~Sanchez-Palencia and D.~I.~Borisov for discussion
of the work and useful remarks. The work is supported by grants of
RFBR (02-01-00693, 02-01-00768) and by the program ''Scientific
Schools'' (1446.2003.1).

\end{document}